# First Report of Susceptibility Status of the Invasive Vector: *Aedes albopictus* to insecticides used in vector control in Morocco

Souhail Aboulfadl[1,2*], Fouad Mellouki[1], Btissam Ameur[3] & Chafika Faraj[2]

*1- Hassan II University of Casablanca, Faculty of Sciences and Technologies Mohammedia, Microbiology, Hygiene and Bioactive Molecules team LVMQB / ETB, BP28806, Morocco.*

*2- National Institute of Hygiene, Medical Entomology Laboratory of Rabat, BP769, Morocco.*

*3- Epidemiology and Disease Control Directorate, Anti-Vector Control Service, BP335, Morocco.*

chafikaf@gmail.com ; $ID_1$: https://orcid.org/0000-0002-8229-3315
fouad.mellouki@univh2c.ma ; $ID_2$: https://orcid.org/0000-0002-1949-9628
btissama@gmail.com ; $ID_3$: https://orcid.org/0000-0002-3816-7741
*Corresponding author: Souhail Aboulfadl, email : souhail.aboulfadl@usmba.ac.ma; $ID_3$: https://orcid.org/0000-0001-9699-5337

**Abstract:**

*Aedes albopictus* has been newly recorded in Agdal district at Rabat in Morocco. The establishment of this invasive mosquito could affect the public health by causing serious epidemics despite of its high nuisance in urban and sub-urban areas. Vector control is mainly based on environmental management but chemical insecticides can be used to reduce adult mosquito densities during peak periods. However, the level of susceptibility of this mosquito to insecticides has not been studied yet in Morocco.

This paper reports the results of the first study conducted to monitor the insecticide resistance of adult and larva *Ae. albopictus* to the insecticides currently used in the vector control. The study was carried out during May-June 2018 at Rabat from the north-west of the country. Adult susceptibility tests were performed following the WHO test procedures. One organochlorate (DDT 4%), one pyrethroids (cyfluthrin 0.15%), one carbamate (bendiocarb 0.1%) and one organophosphate (fenithrothion 1%) were tested at diagnostic doses (DD). The five-fold DD of bendiocarb were also used to yield information on the intensity of mosquito adult resistance. The results of the performed susceptibility bioassay showed that the vector is susceptible to cyfluthrin and resistant to DDT, bendiocarb and fenithrothion.

Larval bioassays to temephos were conducted according to WHO standard operating protocol to establish the dose-mortality relationship and deduct the LC50 and LC90 then resistance ratios. We show that larval populations of *Ae. albopictus* are still sensible to this insecticide.

This information could help policy-makers to plan insecticide resistance management.

**Introduction:**

*Aedes albopictus* is an invasive mosquito called also *Stegomyia albopicta* (Diptera: Culicidae) originally from the south-eastern part of the Asian continent (Petrić et al. 2017). This mosquito has proved an aggressive human and day biting behavior in addition to its high plasticity, thing that has allowed it to colonize a wide range of breeding sites worldwide. It has been established in Europe, in the Mediterranean region more particularly, since the 1990s, and after that it spread more widely to Africa then to the Americas and now occurs in all inhabitable continents (Marcombe et al. 2014). Two well-known ways of species spreading has been reported repeatedly which are eggs and adult's introduction in ornamental plants and used tires and through vehicle bodies respectively. In plus, climatic and environmental changes as well as the continuous globalization facilitate the spread of this invasive mosquito. In Morocco *Ae. albopictus* was first reported in 2015 in the garden of a private dwelling in the Agdal district in the capital Rabat (Bennouna et al., 2016). It was subsequently found in other sites in the same Agdal district and in the neighboring Souissi district (Faraj et al. 2018). The latter authors suggested that quite possibly Agdal district represent the entry point of this species in Morocco.

Following this notification, local authorities regularly conduct mosquito control campaigns against adults using pyrethroids in public spaces. They also carry out awareness campaigns for the benefit of the inhabitants of the district to incite them to eliminate or treat indoor breeding sites.

Unfortunately, the levels and distribution of *Ae. albopictus* susceptibility to insecticides has not been studied in Morocco. Thus, to improve control measures and to provide a rational framework for choosing the suited insecticide, this study has been undertaken. It aimed to investigate and to give basic data about the susceptibility status of this species to insecticides currently used in vector control.

**Materials & methods:**

<u>Mosquito sampling, rearing and identification:</u>

Mosquitoes used in our tests come from eggs collected during May-June 2018 at Rabat. Five ovitraps less than 500 m apart were placed on a street of the Agdal neighborhood in Rabat

(33°59'20.900 N, 6°51007.900W). Ovitraps were checked for eggs once a week and were brought back to the insectary. After hatching, larvae were supplied every 2 days with a yeast tablet dissolved in 1L of dechlorinated tap water. All immature stages were reared at 25 ± 2 °C. Emerging adults were collected by mouth aspirator, transferred to netting cages (30 x 30 x 30 cm) and maintained at 28±1 °C, 80% relative humidity. Adults designated for breeding were fed by 5% agarose solution and the ones designated for bioassays were kept unfed. Larvae were identified under microscope using the taxonomic key : the Mosquitoes of Europe Software (MousEurope) (Schaffner et al. 2001).

Larval Bioassays:

We studied the state of susceptibility of *Ae. albopictus* to temephos, this insecticide belonging to organophosphate family is authorized in the fight against larvae in indoor breeding sites in Morocco. Bioassays were carried out following WHO standard procedures to establish the dose-mortality relationship and to calculate lethal concentrations $LC^{50}$ and $LC^{95}$ (concentrations involving, resp., the death of 50% and 95% of the tested population) (WHO, 2017). Tests were carried out on the third and fourth instar larvae. Ranges of 5 concentrations of temephos and control were prepared to determine the $LC^{50}$ and $LC^{95}$ for the tested population. For each dilution, 5 replicates were done, with 25 larvae each. Larvae were placed in 99 mL of water; 1 mL of adequate concentration of temephos was then added. After 24 hours of exposition at ambient temperature (21-22 °C) without feeding, alive and dead larvae were counted.

Adult Bioassays:

Bioassays were performed following WHO protocols (WHO, 2016b). Unfed *Ae. albopictus* females of 3-5 days old were tested against four insecticides belonging to four classes as follows: cyfluthrin 0.15% (pyrethroid), DDT 4% (organochloride), bendiocarb 0.1% (carbamate) and fenithrothion 1% (organophosphate). We determined the resistance intensity by using 5× DD of bendiocarb (0.5%). In each experiment 20 unfed females were placed in OMS tubes containing the insecticide treated papers with 5 replicates per test. Two controls were used for each test containing insecticide untreated papers and 20 non-fed females each (Aboulfadl et al. 2020). Tested mosquitoes were then provided with 10% sugar solution and maintained at 28 ± 2 °C and 80 ± 10% RH, and then overall mortality was observed after 24 hours. The knock-down time ($KdT_{50}$ and $KdT_{90}$ indicating 50% and 90% of knocked-down tested populations respectively) was calculated for pyrethroids and organochlorine exposed mosquitoes using WinDL32 software (Giner et al., 1999).

Data analysis:

We performed larval bioassay data using Log-probit analysis software, WinDL version 2.0, to fit a linear regression between the log of the insecticide concentration and the probit of mortality and to estimate thelethal concentrations (LC50 and LC90) with their95% confidence intervals (CI) (Giner et al.1999). Larval mortality rates were corrected using Abbott's formula (Abbott 1925) in case of control mortality is greater than 5% but less than 20%.

Based on the LC90 obtained from larval bioassays,we calculated resistance ratios (RR) by dividing values for the tested strains by those of the susceptible strain. Susceptible strain data were obtained from Ishak et al. (2015) study.

We considered a mosquito population susceptible when RR90 was less than 2, potentially resistant when RR90 was between 2 and 5, and resistant when RR90 was over 5 (Kamgang et al. 2011). Resistance is considered high when calculated RR90 values > 10, moderate when 5_RR90 _ 10 and low when RR90 <5 (Mazzarri and Georghiou 1995).

Concerning adult bioassay, the WHO criteria were adopted for distinguishing between resistance/susceptibility status and resistance intensity of the tested mosquito populations (WHO, 2016b). When more than 98% mortality at DD was observed, the population was considered "susceptible", less than 90%; the population was considered "resistant". Mortality of 98–100% at the 5xDD indicates low resistance intensity. Mortality of <98% but ≥ 98% at the 10×DD, resistance intensity is moderate and when mortality <98% at the 10× DD, resistance intensity is high.

**Results:**

Susceptibility of *Aedes albopictus* larvae

Our results showed that the *Agdal* strain presented an $LC_{50}$ which is equal to 0.0065 mg/L very close to the $LC_{50}$ of the sensitive strain VCRU from Malaysia ($LC_{50}$= 0.006), and an $LC_{90}$ which is equal to 0.010 mg/L with a very low resistance ratio (RR) of 1.08 compared to the susceptible strain VCRU (Table 1).

**Table 1:** Susceptibility of *Aedes albopictus* larvae to temephos in the Rabat region.

| Concentrations (mg/L) | $CL^{50}$ (mg/L) [IC] | $CL^{90}$ (mg/L) [IC] | $CL^{100}$ Observed | $RR^{50}$ | $CL^{50}$ VCRU strain | Statut |
|---|---|---|---|---|---|---|
| 0.005 ; 0.007 ; 0.01 ; 0.017 ; 0.02 | 0.0065 [0.0018-0.023] | 0.010 [0.006-0.018] | 0.017 | 1.08 | 0.006 | S |

Susceptibility of *Aedes albopictus* adult

Results showed that adult populations of *Ae albopictus* studied were resistant to the OC, OP and CX families, but completely susceptible to the PYRs. The lowest observed mortality was against DDT4% and equal to 67%, while the highest mortality was against bendiocarb 0.1% and fenithrothion 1% and equal to 92%. The populations exposed to cyfluthrin 0.15% showed complete susceptibility with 100% mortality. In addition, we calculated the time of knockdown induced by DDT and cyfluthrin and we showed that the knockdown time $KdT_{50}$ and $KdT_{90}$ are variable depending on the active ingredient of the insecticide used. The $KdT_{50}$ of DDT 4% has a value of 142 which indicates strong resistance in the tested population. Regarding the $KdT_{50}$ of cyfluthrin 0.15%, its value ( 16.2) is considerably lower than that of DDT. The resistance intensity of *Ae. albopictus* adults to carbamates was low with a mortality rate of 100% after being exposed to bendiocarb 0.5%.

**Table2:** Susceptibility bioassays results of the of adult *Ae. albopictus* mosquitoes to organochlorines, pyrethroids, carbamates and organophosphates respectively in the region of Rabat.

| Insecticides | Rabat | | | | |
|---|---|---|---|---|---|
| | N | TKd50 [IC] | TKd90 [IC] | %Mr | Statut |
| DDT 4% | 102 | 142 [15.8- 1274] | 378 [12.4- 11495] | 67 | R |
| Cyfluthrin 0.15% | 100 | 16.2 [11.9- 20] | 22.5 [18.5- 35.3] | 100 | S |
| | N | %Mr | | Statut | |
| Bendiocarb 0.1% | 92 | 92 | | R | |
| Bendiocarb 0.5% | 100 | 100 | | RI : LR | |
| Fenithrothion 1% | 97 | 92 | | R | |

Legend : N : Number of mosquitoes ; $KdT_{50}$ : Time of which 50% of the tested population suffers a Knock down; $KdT_{90}$ : Time of which 90% of the tested population suffers a Knock down; %Mr : Mortality percentage ; RI : Resistance intensity; LR : Low resistance; IC: 95% confidence interval.

**Discussion:**

Our study is the first in Morocco that provides basic data on the larval and adult susceptibility status of the tiger mosquito *Ae. albopictus*. To our knowledge, there are no data available on the susceptibility of this species in Mediterranean Africa either. The objective of this study was to characterize and quantify the intensity of the Moroccan population resistance to the insecticides currently used in vector control strategies in the country. It showed that the larvae are fully susceptible to temephos whereas the adults are susceptible to cyfluthrin but resistant to DDT, bendiocarb and fenitrothion. The intensity of resistance was low to bendiocarb.

In Morocco, the main insecticides recommended to control mosquitoes are temephos, which is used against larvae, and pyrethroids, which are used against adults. For these two insecticides, we observed a normal susceptibility of the tested population. These insecticides can therefore continue to be used by vector control services while continuing to regularly monitor the evolution of vector susceptibility to these products. The use of temephos is obviously only recommended if the physical control methods which aims to eliminate the breeding sites, are not appropriate.

Early physical and environmental actions will likely prevent the development of resistance within larval populations and ensure successful and effective use of this insecticide for a long time.Our results agree with many previous ones who showed the susceptibility of *Ae. albopictus* against temephos in Asia, America and Europe (Dalla et al. 1994; Lee et al. 1998; Rohani et al. 2001; Marcombe et al. 2014; Mohiddin et al. 2015; Sivan et al. 2015 and Bharati and Saha, 2017). However, larval populations have shown a high resistance to temephos in countries where they are subjected to prolonged exposure in the field (Ranson et al. 2010; Vontas et al. 2012; Gómez et al. 2011 Sivan et al. 2015; Arslan et al. 2016; Rath et al. 2018; Su et al. 2019; Li et al. 2020).

One of the most important findings of our study is the low mortality, and therefore high resistance, of the *Ae. albopictus* adults to DDT although this insecticide has not been used in vector control for decades. Resistance to this insecticide has been previously described in previous studies (Singh et al. 2013; Arslan et al., 2016; Li et al. 2020; Wan-Norafikah et al. 2021).

The two most widely recognized mechanisms of insecticide resistance are voltage-gated channel modification (knockdown resistance mutation; *kdr*) and the overproduction of detoxification enzymes (Auteri et al. 2018). Even though DDT target sodium channels, resistance mechanisms are not specific to this particular insecticide family but to pyrethroid insecticide also. However, no resistance to cyfluthrin was detected in this species, which confirms that the resistance observed for DDT is totally metabolic.

The susceptibility of *Ae. albopictus* to cyfluthrin is consistent with previous studies reporting relative susceptibility of this species to pyrethroids in India, Malaysia, Thailand, Cameroon, Greece and Italy (Vontas et al. 2012). However, moderate to high resistance was observed in *Ae. albopictus* in several regions of the world (Ishak et al. 2015; Li et al. 2020; Gómez et al. 2011; Su et al. 2019; Arslan et al. 2016; Wan-Norafikah et al. 2021). Resistance to bendiocarb and fenitrothion has been observed in other countries (Sivan et al. 2015). However, no resistance to bendiocarb was detected in this species in Cameron (Yougang et al. 2020). *Ae.albopictus* showed no resistance to fenitrothion in studies carried out in India (Ranson et al. 2010), Thailand (Chuaycharoensuk et al. 2011) and Cameroon (Kamgang et al. 2011).

In our study, the susceptibility observed in larval populations against temephos and adult populations against cyfluthrin is mainly due to the absence of insecticide pressure and the scarcity or absence of resistant individuals.On the other hand, the resistance expressed in the adult phenotype of *Ae. albopictus* from Rabat against bendiocarb and fenitrothionis nascent and requires early management. Because of the absence of any chemical control by public hygiene services at the time when we carried out the tests, this resistance could not be induced, by the selection pressure triggered by localities. Thus, it is most likely due to a resistant genetic ancestor whose descent was transported to Morocco. It may be due to many mechanisms, in particular metabolic, cuticular or by target site modification. The synergist PBO4% results cannot be interpreted because the mortality percentage of mosquitoes exposed to bendiocarb 0.1% only was greater than 90%.Other additional studies must be carried out to demonstrate the resistance mechanisms in adult populations.

It is quite clear that the *Agdal* strain does not show cross-resistance between PYR and DDT, which means that resistance to DDT is most likely not linked to the kdr gene or to a common enzyme. So, in order to maintain the current static state of susceptibility of larvae and adults to temephos and pyrethroids respectively, it requires strong government involvement to be careful with insecticide application and provide more alternative methods of control.

This is the first report of the susceptibility status of *Ae. albopictus* from Morocco, that gives basic data that will help the authorities to monitor the species in order to be able to effectively manage the resistance that has appeared within individuals and therefore maintain the durability and effectiveness of chemical insecticides, or to also act profitably in the event of an epidemic or possible epizootics.

**Conclusion:**

The presence of *Ae. albopictus* mosquito in the Moroccan capital threats the public health by possible transmission of arboviruses and will induce an economic burdenwhether it was spread across the country. Our work represents the first study in Mediterranean Africa on the susceptibility status of the Moroccan populations to the four classes of chemical insecticides most used in public health. Our results have shown susceptibility in *Ae. albopictus* larvae to temephos and resistance in adult population against carbamates, organochlorides and organophosphates but a susceptibility against pyrethroids.Despite of the susceptibility shown against pyrethroids and temephos, the surveillance of the susceptibility status of this species and its distribution should be a public health priority for the country to avoid the resistance development or the mosquito invasion in other cities.